\documentstyle[preprint,aps]{revtex}
\tightenlines
\begin{document}
\draft
\title{THEORETICAL CONTINUOUS EQUATION DERIVED FROM THE MICROSCOPIC DYNAMICS
FOR GROWING INTERFACES IN QUENCHED MEDIA}
\author{L. A. Braunstein \cite{mimail}, R. C. Buceta, C. D. Archubi}
\address{Departamento de F\'{\i}sica, Facultad de Ciencias Exactas y
Naturales, Universidad Nacional de Mar del Plata, Funes 3350, 7600
Mar del Plata, Argentina}
\author{G. Costanza}
\address{{Departamento de F\'{\i}sica, Universidad Nacional de San Luis,
Chacabuco 917, 5700 San Luis, Argentina}}

\maketitle

\begin{abstract}
We present an analytical continuous equation for the Tang and
Leschhorn model [Phys. Rev A {\bf 45}, R8309 (1992)] derived from
his microscopic rules using a regularization procedure. As well in
this approach the nonlinear term $(\nabla h)^2$ arises naturally
from the microscopic dynamics even if the continuous equation is
not the Kardar-Parisi-Zhang equation [Phys. Rev. Lett. {\bf 56},
889 (1986)] with quenched noise (QKPZ). Our equation looks like a
QKPZ but with multiplicative quenched and thermal noise. The
numerical integration of our equation reproduce the scaling
exponents of the roughness of this directed percolation depinning
model.

\end{abstract}

The investigation of rough surfaces and interfaces has attracted
much attention for decades due to is importance in many fields,
such as the motion of liquids in porous media, growth of bacterial
colonies, crystal growth, etc. When a fluid wet a porous medium, a
nonequilibrium self-affine rough interface is generated. The
interface has been characterized through scaling of the
interfacial width $w=\langle[h_i-\langle
h_i\rangle]^2\rangle^{1/2}$ with time $t$ and lateral size $L$.
The result is the determination of two exponents, $\beta$ and
$\alpha$ called dynamical and roughness exponents respectively.
The interfacial width $w\sim L^\alpha$ for $t\gg t^*$ and $w\sim
t^\beta$ for $t\ll t^*$, where $t^*=L^{\alpha/\beta}$ is the
crossover time between this two regimes. Much effort has been done
to understand the leading mechanisms of these processes and to try
to explain how the dynamics affects the scaling exponents
\cite{Family}. The formation of interfaces is determinated by
several factors, it is very difficult to discriminate
theoretically all of them. The knowledge of the dynamical
nonlinearities, the disorder of the media, and the theoretical
model representing experimental results are difficult to overcome
due the complex nature of the growth. The disorder affects the
motion of the interface and leads to its roughness. Two main kinds
of disorder have been proposed: the ``annealed'' noise that
depends only of time and the ``quenched'' disorder due to the
inhomogeneity of the media where the moving phase is propagating.
The discrete models provided an useful approach to obtain the
exponents that allows its classification in universality classes.
By extensively studying these models, one can obtain the scaling
behaviors and the corresponding universality classes, and then
associate the continuum stochastic equations with the given
discrete growth models.

The most used method of establishing the correspondence between a
continuum growth equation and a discrete model, is to numerically
simulate the model and compare the obtained scaling exponents with
those of the corresponding continuum equation. In this context
attempts are being made to classify quenched disorder models in
terms of universality classes based on equation of motion such as
\begin{equation}
 \frac{\partial h({\bf x},t) }{\partial t}=\;{\cal{F}}+\nu \nabla ^2 h +
\frac{\lambda}{2} (\nabla h)^2 +\xi({\bf x},h) +\eta({\bf x},t)
\label{qkpz}\;,
\end{equation}
where ${\cal{F}}$ is the driving force responsible of the advance
of the interface, $\xi({\bf x},h)$ is the quenched disorder or
pinning forces and $\eta({\bf x},t)$ is the thermal noise. The
noises are white. Eq.~(\ref{qkpz}) is Kardar-Parisi-Zhang equation
\cite{Kardar} with quenched noise (QKPZ). When $\lambda \to 0$ the
quenched Edward-Wilkinson equation (QEW) \cite{Edwards} is
recovered. In absence of quenched noise his thermal versions are
recovered, named KPZ and EW respectively. Much effort has been
done in order to classify discrete models and experiments in
universality classes depending on the value of the coefficient
$\lambda$ associated with the nonlinearity of the QKPZ. Numerical
\cite{Amaral,Reka} studies indicate $\lambda$ is relevant at the
depinning transition for discrete models in anisotropic media.
These results only shows that the nonlinear term exist but they do
not confirm that these models are represented by the QKPZ. However
the exponents obtained by numerical simulation of
Eq.~(\ref{qkpz}), without thermal noise \cite{Lesch}, agree very
good with those of the model in anisotropic media.

A powerful method of establishing the correspondence between a
continuum growth equation and a discrete model is to derive the
continuum equation from a given discrete model analytically. Among
them, a systematic method proposed by Vvedensky {\it et al.}
\cite{Vvedensky}, where the continuum equations can be constructed
directly from the growth rules of the discrete model based on the
master-equation description, has been applied to the derivation of
growth equations for some discrete models
\cite{Vvedensky,Park,Park2} with thermal noise. This method has
proved to be useful to derive continuous equation from the
microscopic ones with the advantage that the sources of the terms
of the Langevin equation can be identified and their magnitude
related to the microscopic dynamics. While the method described by
theses authors can be applied to general case it is more easy to
achieve the same results using a microscopic equation based on
rules for the evolution of the height. The derivation of
continuous equations from discrete models is an interesting
subject which has not been addressed in the context of growth in
presence of quenched media.

The aim of this work is to obtain the continuous equation from the
microscopic dynamics of a model with quenched noise in order to
establish if it is related in some way with the QKPZ equation. The
main goal of our paper is to have obtained analytically the
differential equation which describe the dynamics of the Tang and
and Leschhorn (TL) model \cite{Tang}. To our knowledge this is the
first time that a Langevin equation is obtained from the
microscopic dynamics in quenched media.

As we shall show below the dynamics of the height is strongly
affected by a {\sl multiplicative quenched noise}. We chose this
model because it presents the principal features of some
experiments like the imbibition of a viscous fluid in a porous
media driven by capillary forces \cite{Buldyrev,Horvath}. In the
TL model, the interface growth takes place in a square lattice of
edge $L$ with cells of size $a$ that represents the mean size of a
pore. For a given applied pressure $p>0$ , we can divide the cells
into two groups, those with $g({\bf r}) \le p$ (free or active
cells), and those with $g({\bf r}) > p$ (blocked or inactive
cells). Denoting by $q$ the density of inactive cells on the
lattice we have $q=1-p$ for $0< p<1$ and $q=0$ for $p\ge 1$. In
this model the critical pressure is $p_c=0.461$. Periodic boundary
conditions are used. We consider the evolution of the height of
the $i$-th site in this model. Let us denote by $h_i(t)$ the
height of the $i$-th generic site at time $t$. The set
$\{h_i,i=1,\dots,N\}$ where $N=L/a$, defines the interface between
wet and dry cells. Given a site, chosen between $N$, say the site
$j$, the height in the site $i$ is increased by $a$ with
probability (i) $1$ if $j=i\pm 1$ and $h_{i\pm 1} \ge h_i+ 2a$ and
$h_i< h_{i\pm 2}$, (ii) $1/2$ if $j=i\pm 1$ and $h_{i\pm 1} \ge
h_i + 2a$ and $h_i = h_{i\pm 2}$, (iii) $1$ if $j=i$ and $h_i<
\min(h_{i-1},h_{i+1}) + 2 a$ and $F_i(h_i+a)=1$. Otherwise no
growth happens. $F_i(h_i+a)=\Theta(p-g_i(h_i+a))$ is called the
activity function \cite{Brauns1} and $\Theta(x)$ is the unit step
function defined as $\Theta(x)=1$ for $x\ge 0$ and equals to $0$
otherwise, $p$ is the microscopic driving force and $g_i(h_i+a)$
is the quenched noise just above the interface distributed in the
interval $[0,1]$. Notice that the activity function $F$ is the
competition between the driving force and the quenched noise, so
$F$ is also a ``noise''. Provided that the system size is large
and that the intrinsic fluctuations are not too large
\cite{Vvedensky}, the evolution equation for the height in a site
$i$, in a short lapse $\tau$, is
\begin{equation}
\frac{\partial h_i}{\partial t} = \frac{a}{\tau} G_{i} +
\eta_i\label{1}\;,
\end{equation}
where $\tau$ is the mean lapse between successive election of any
site and $G_{i}$ \cite{Brauns1} are the microscopic growing rules
for the evolution of the height at this site due that a site $j$
is chosen at time $t$. Here $\eta_i$ is a Gaussian ``thermal''
noise with zero mean and covariance
\begin{equation}
\langle \eta_i(t)\eta_j(t')\rangle = \frac{a^2}{\tau}\;
G_i\;\delta_{ij}\;\delta(t-t')\;.
\end{equation}
Notice that in the notation of \cite{Vvedensky} the transition
rate from from a configuration $H$ to another $H^{'}$ is $
W(H,H^{'})=\frac{1}{\tau} \sum_{k} \delta(h^{'}_k,h_k+a) G_k
\prod_{j\neq k} \delta(h^{'}_j,h_j)$. So the first moment is $
\frac{1}{\tau}\;\sum_{H^{'}} (h^{'}_k -h_k) W(H,H^{'})
=\frac{a}{\tau} G_k$. As a consequence of the fact that subsequent
configurations differ only in the height at one site all the
moments are diagonal and proportional to the first moment
\cite{Park2}.

For this model \cite{Brauns1,Brauns2},
\begin{eqnarray}
G_{i}(h_{i-1},h_i,h_{i+1})=W_{i+1}+
W_{i-1}+F_i(|h_i|+1)\;W_i\label{Gi}\;,
\end{eqnarray}
where $|h_i|\equiv [\frac{h_i}{a}]$ denotes the integer part of
$h_i$ in units of $a$. This definition is meaningless if $a$ is
taken as one, as in the discrete model. We shall show below that
in the continuous limit, it means that $F$ is taken as a constant
inside a pore. In Eq.~(\ref{Gi}),
\begin{equation}
\begin{array}{l}
W_{i\pm 1}=\case{1}{2}\;[1-\Theta(H_{i\pm 2}^i)+\Theta(H^{i\pm
2}_i)]\;\Theta(H^{i\pm 1}_i-2)\;,\\\hspace{.4cm}W_i =
1-\Theta(U_i-2)\;,
\end{array}\label{wi}
\end{equation}
where $H_r^s=(h_s-h_r)/a$ and
$U_{i}=\frac{1}{a}[h_i-\min(h_{i+1},h_{i-1})]$. Notice that all
the heights are in units of $a$ in order to keep the arguments of
the step function without units. For $W_{i \pm 1}$ \cite{Brauns1}
the $\delta$ Kronecker function  has been taken as
\begin{equation}
\delta(x,y)=\Theta(x-y)+\Theta(y-x)-1 \label{delta}\;.
\end{equation}
Using the fact that $ \min(x,y)= \frac{1}{2}
\{(x+y)-(x-y)[\Theta(x-y)-\Theta(y-x)] \}$ and with a more compact
notation
\begin{equation}
U_{i}=\case{1}{2}\{H_{i+1}^{i}+ H_{i-1}^{i}+H^{i+1}_{i-1} \;
[\Theta(H^{i+1}_{i-1}) -\Theta(H^{i-1}_{i+1})]\}\label{u}\;.
\end{equation}

The representation of the step function can be expanded as
$\Theta(x)=\sum_{k=0}^{\infty} c_k x^k$ providing that $x$ is
smooth. Our focus is on properties of the surface on large length
scales so we kept the expansion of the step function to first
order in his argument. The best choice for the representation of
the $\Theta$ function is the shifted hyperbolic tangent
\cite{Park2}, defined as $\Theta(x)=\{1+ \tanh [C (x +b)]\}/2$,
where $b$ is the shift and $C$ is a parameter that allows to
recover the $\Theta$ in the limit $C \to \infty$. We choose
$b=1/2$. The reason of our choices is that it allows us to define
the $\delta$ function as Eq.~(\ref{delta}). The coefficients
fulfill
\begin{equation}
c_0=\frac{1}{2}\;[1+ \tanh
(\frac{C}{2})]\hspace{1cm}\mbox{and}\hspace{1cm} c_1=\frac{C}{2}\;
\cosh^{-2}(\frac{C}{2})\label{c0}\;.
\end{equation}
We shall show bellow the qualitative information that can be
obtained from these coefficients assuming this regularization
scheme. The next step is to regularize the height defining an
interpolating function. This is done by expanding the height
$h_{i+\ell}\equiv h(x_i+x_\ell)$ around $x_i=i a$. Retaining only
the leading terms in the expansion, the adimensional difference of
heights is
\begin{equation}
H^{i+\ell}_{i+m} = (\ell-m)\;\partial_x h\rfloor_{x_i}  +
\case{1}{2}(\ell^2-m^2)\;
\partial_x^{2} h\rfloor_{x_i}\;a
+{\cal{O}}(a^2)\label{H}\;,
\end{equation}
where $\partial_x^j h=\partial^jh/\partial x^j$.

Notice that in any discrete model there is in principle an
infinite number of nonlinearities, but at long wavelengths the
higher order derivatives can be neglected using scaling arguments,
since one expect affine interfaces over a long range of scales,
and then one is usually concerned with the form of the relevant
terms.

Replacing Eq.~(\ref{H}) in Eqs.~(\ref{wi}) and (\ref{u}), using
the expansion of the step function and retaining the leading terms
to order ${\cal{O}}(a)$, Eq.~(\ref{1}) can be written as
\begin{equation}
\frac{\partial h(x_i,t)}{\partial
t}=\frac{a}{\tau}\;\left[W(x_i+a)+
W(x_i-a)+W(x_i)\;F(x_i,|h(x_i)|+1)\right]
+\eta(x_i,t)\label{dhdt}\;,
\end{equation}
with
\begin{eqnarray}
W(x+a)+W(x-a)&=&(c_0- 2c_1) +4 c_1^2\;(\partial_x h)^2+ a
c_1\left[\case{1}{2}+4(c_0-2c_1)\right]\;
\partial_x^{2}h\label{difu}\;,
\\ W(x)&=&1-(c_0-2 c_1) - 4 c_1^2 \;(\partial_x h)^2 + \case{1}{2} a c_1
\;\partial_x^2 h\label{sustrato}\;.
\end{eqnarray}
Notice that the argument of $F=\Theta(p-g(x_i,|h(x_i)|+1))$ is not
smooth, so his expansion is meaningless. For the activity function
$F$ we define an interpolation function
\begin{equation}
\widetilde{F}(x_i,h(x_i))=F(x_i,|h_i|)+\frac{\delta h}{a}
[F(x_i,|h_i|+1) - F(x_i,|h_i|)]+ {\cal{O}} (\delta h^2)\;,
\label{interF}
\end{equation}
with $0 \leq \delta h \leq a$ that measures the departure of the
height from the low pore. Then $\widetilde{F}$ is a smooth
function taking continuous values in the interval $[0,1]$. With
this definition we ensure that the characteristic size of the
correlation between pores is of the order of the pore size. In
real materials there always exist a typical size of the
inhomogeneities in the disordered media which plays the role of
the lattice constant $a$.

The final step is a coarse-grained spatial average of the
variables in order to obtain smooth continuous functions at a
macroscopic level. In this way we obtain the stochastic continuous
equation for this model,
\begin{equation}
\frac{\partial h}{\partial t}=\mu(\widetilde{F})+
\nu(\widetilde{F})\;\partial^2_x h + \lambda(\widetilde{F})\;
(\partial_x h)^2 +\eta(x,t)\;,\label{bbac}
\end{equation}
where
\begin{eqnarray}
\mu(\widetilde{F})&=&[(c_0-2 c_1) (1-\widetilde{F})
+\widetilde{F}] \,\frac{a}{\tau}\;,\label{df}\\
\nu(\widetilde{F})&=&c_1[\case{1}{2}\,(1+\widetilde{F})+4 (c_0-2
c_1)]\,\frac{a^2}{\tau}\;,\\
\lambda(\widetilde{F})&=&4\,c_1^2\,(1-\widetilde{F})\,\frac{a}{\tau}\;.
\end{eqnarray}
and $\widetilde{F} \equiv F(x,h)$ as was defined in
Eq.~(\ref{interF}). Notice that $\mu(\widetilde{F})$ is now the
effective competition between the driving force and the quenched
noise.

Equation~(\ref{bbac}) shows that the nonlinearity {\sl arises
naturally} as a consequence of the microscopic model.

This explains the previous numerical results obtained by Amaral
{\sl et al.} \cite{Amaral}, that studied the effects of an
effective coefficient $\lambda_{eff}$ from a tilted interface
showing that the nonlinear term must exist. Our result is also in
agreement with those of R\'eka {\sl et al.} \cite{Reka} that
obtained numerically a parabolic shape of the local velocity as
function of the gradient for the DPD model near above the
criticality for different reduced forces $(p/p_c-1)$. Now, let us
consider for a moment the restrictions we have imposed over the
coefficients $c_0$ and $c_1$ in the expansion of the
representation of the $\Theta$ function. When the conditions
(\ref{c0}) are satisfied, the coefficient $\nu$ is always
positive. The coefficient $\lambda$ is greater or equal to zero
independently of the representation of the step function. On the
other hand $\mu$ can be positive or negative depending of the
value of $\widetilde{F}$. In Fig.~\ref{uno} we show
$\widetilde{F}$ as function of $C$ for $\lambda=\nu$ (upper curve)
and $\nu=0$ (lower curve). We can distinguish three regions. In
the region $\mu < 0$ and $\lambda > \nu >0$, the function
$\widetilde{F}$ takes values close to zero. Notice that in the
discrete model, as we approach to the critical value $F$ is mostly
zero because the interface gets pinned by long chains of inactive
sites\cite{Brauns1,Brauns2}. In the continuous model near the
criticality $\mu$ mostly becomes negative braking the advance of
the interface and $\widetilde{F}$ is close to zero. Then regions
of small gradients will be halted by the effective driving force
$\mu$. In these case the main responsible of the nonlinearities is
the lateral contribution [see Eq.~(\ref{difu})]. This explain why
this contribution enhances the roughness at the criticality as was
predicted by Braunstein {\sl et al.} \cite{Brauns1}.

Faraway above the criticality the nonlinear term becomes less
relevant. In the limit $p \to 0$ ($p \to 1$) we recover the KPZ
(EW) equation with thermal noise. However the fact that these
limits are recovered is a specific characteristic of this
particular model.

In Figs.~\ref{dos} we show the temporal scaling behavior of the
roughness $w$ obtained from the numerical integration of
Eq.~({\ref{bbac}). At the criticality a slope $\beta=0.67 \pm
0.05$ was obtained. Above the threshold we recover a crossover
between the exponent $\beta= 0.67 \pm 0.05$ and the $\beta_m
\simeq 1/3$ as was obtained by Leschhorn \cite{Lesch} by means of
the numerical integration of the QKPZ equation and by his
automaton version. In Fig.~\ref{tres} we show the scaling behavior
of the correlation function $C_2(r,t)=\langle
[h_{i+r}(t)-h_i(t)]^2\rangle^{1/2}$. The exponent obtained was
$\alpha=0.641 \pm 0.07$ in agreement with the DPD models. The
numerical integration was made in short lattices. The results in
large systems and the details of the integration will be published
elsewhere. Notice that even if the exponents from our equation are
very similar that the one obtained from the QKPZ one, our equation
is very different. The main difference is that the coefficient of
the nonlinear term in our equation is strongly affected by the
local characteristic of the substratum. Another great difference
is that the global velocity of the interface is affected by the
coefficient of the Laplacian term. The study of these differences
requires a great computational effort that is not in the aim of
the present work.

How our results could be used in order to explain the role played
by the disordered media in the experiments? In the experiments the
advancement of the interface is determinated  by the coupled
effect of the random distribution of the capillary sizes, the
surface tension and the local properties of the flow, so it is not
surprising that all these effect give rise to a multiplicative
noise in any evolution equation that intends to represent an
experimental growth with disordered media.

Summarizing, we derive the continuous equation from the
microscopic one for the TL model. In our work the nonlinear term
arises naturally as consequence of the microscopic dynamics. The
numerical integration of our equation reproduces very accurately
the scaling exponents of the roughness. However these results show
that the DPD model is not described by a QKPZ equation with
additive noise even if the exponents of the roughness are very
similar. Our equation also allows to explain that the lateral
growth contribution is the main responsible of the roughness near
the criticality. Finally, we hope that this framework can be used
in other growing models with quenched noise.

\newpage
\begin{figure}
\caption{$\widetilde{F}$ as function of $C$ for
$\mu(\widetilde{F})=0$ (dashed line) and
$\lambda(\widetilde{F})=\nu(\widetilde{F})$ (solid line). In the
first region (I) $\lambda(\widetilde{F}) > \nu(\widetilde{F})>0$
and $\mu(\widetilde{F}) < 0$, in the second region (II)
$\lambda(\widetilde{F}) > \nu(\widetilde{F})>0$ and
$\mu(\widetilde{F}) > 0$ and in the third region (III) $0 <
\lambda(\widetilde{F}) < \nu(\widetilde{F})$ and
$\mu(\widetilde{F}) > 0$ \label{uno}}
\end{figure}
\begin{figure}
\caption{log-log plot of the square roughness $w^2$ vs time for
$C=1.3$. In (a) $p=0.1$, for this value of $C$ the critical
pressure is $p_c \simeq 0.1$. In circles we shows the results
obtained from the numerical integration of Eq.~(\ref{bbac}). The
dashed line is used as a guide and as exponent $2 \beta=1.34$. In
(b) $p=0.3$, the dashed line has slope $2 \beta=1.34$ and the
solid line has slope $2 \beta_m =0.66$. The numerical integration
has been done with $L=1024$ and over $30$ independent
samples.\label{dos}}
\end{figure}
\begin{figure}
\caption{log-log plot of $C_2(r)$ as function of $r$ for $p=0.1$
and $C=1.3$. The dashed line that is used as a guide has slope $2
\alpha=1.28$\label{tres}}
\end{figure}
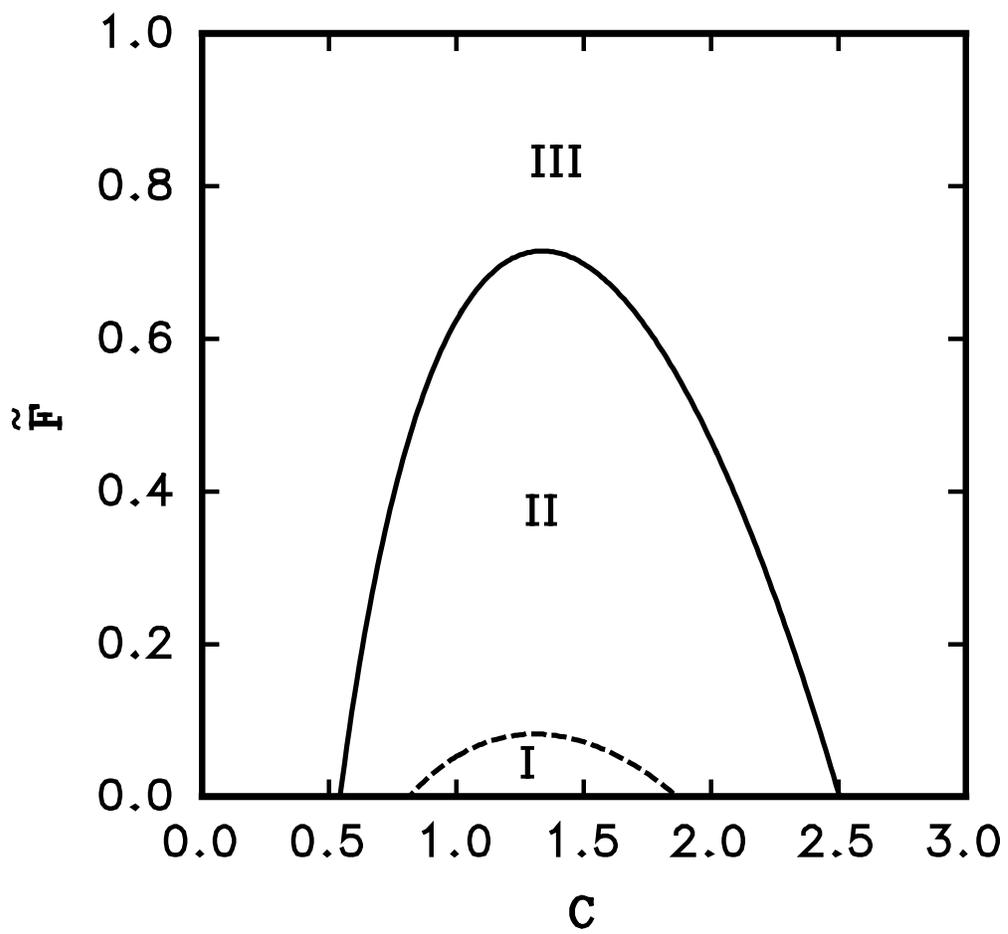
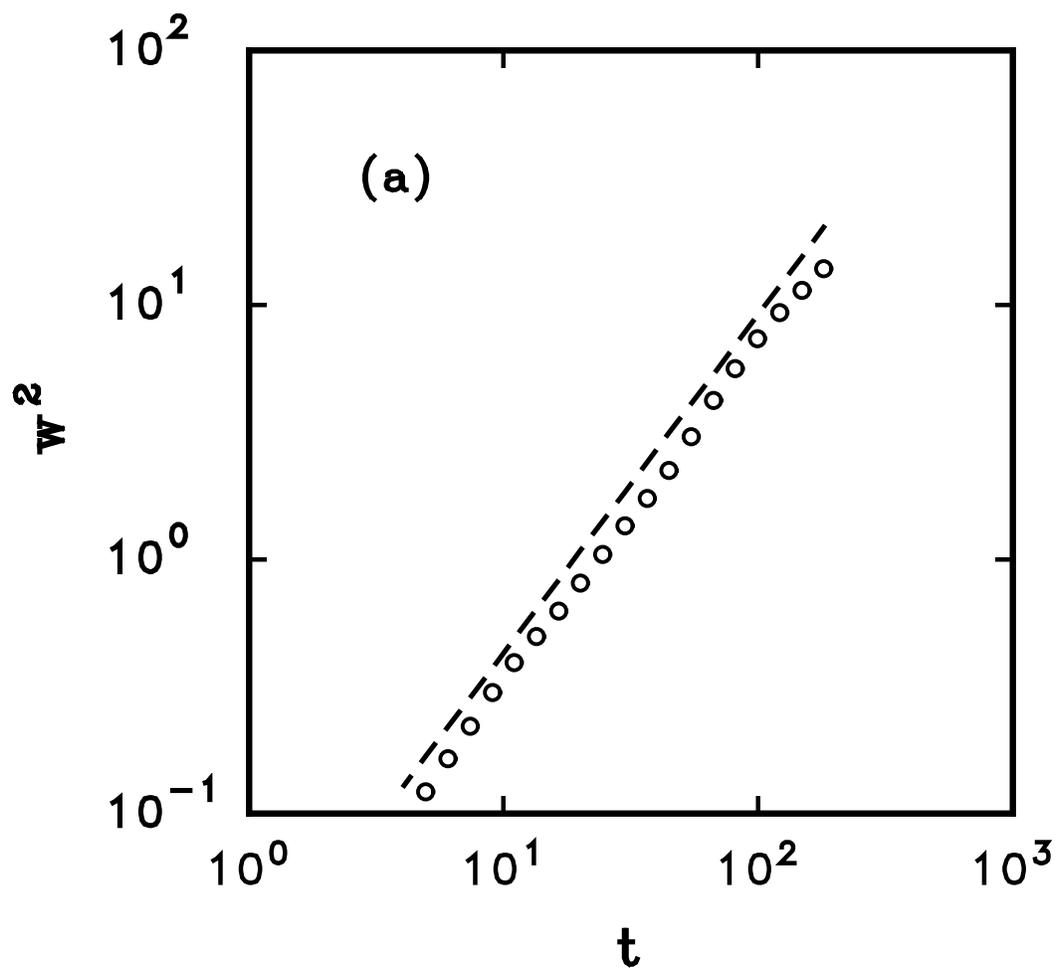
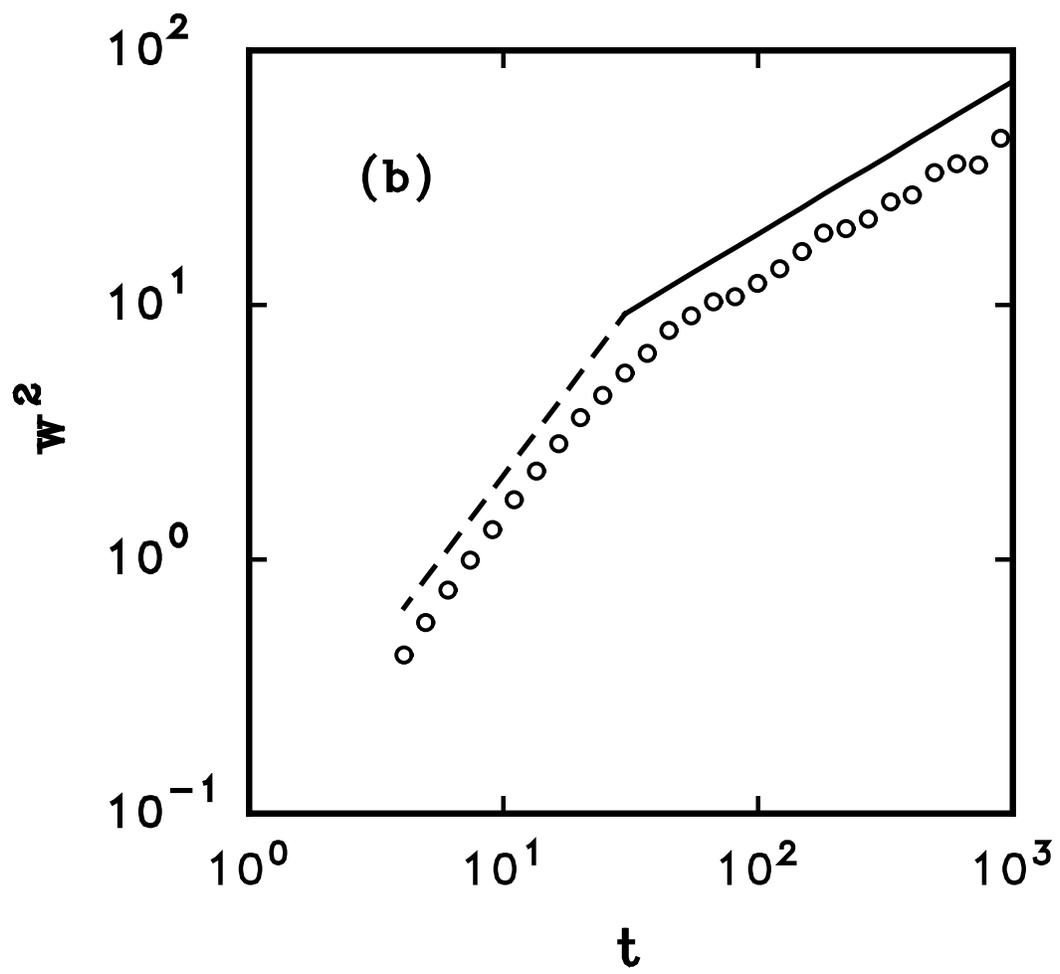
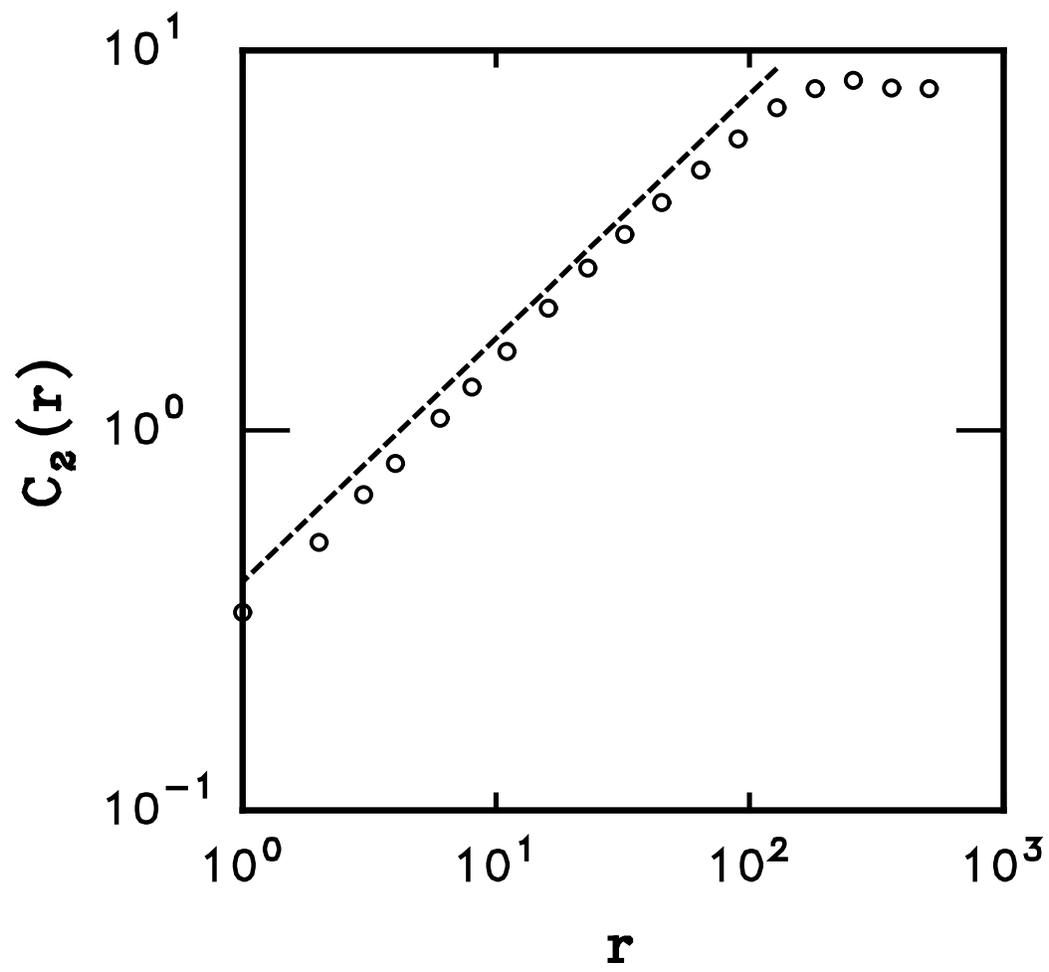
\end{document}